\documentstyle[sprocl]{article}

\def\Journal#1#2#3#4{{#1} {\bf #2}, #3 (#4)}

\def\NPB{{\em Nucl. Phys.} B}
\def\PLB{{\em Phys. Lett.}  B}
\def\PRL{\em Phys. Rev. Lett.}
\def\PRD{{\em Phys. Rev.} D}

\def\be{\begin{equation}}
\def\ee{\end{equation}}
\def\bea{\begin{eqnarray}}
\def\eea{\end{eqnarray}}

\begin{document}

\title{Modulino as natural candidate for a sterile neutrino\footnote{Talk given at the workshop on Phenomenological Aspects of Superstring Theory, 2-4 October 1997, Trieste, Italy.} }

\author{Karim Benakli}

\address{Phys. Dept.  Texas A \& M, College Station \\TX
77843, USA.\\E-mail: karim@chaos.tamu.edu}

\maketitle\abstracts{ We discuss the possible  generation  of R-parity 
violating 
bilinear terms $ \mu_i\bar L_i H_2$ in both cases of gravity  and gauge
mediated  supersymmetry breaking.  Some phenomenological
aspects are reviewed.  In particular for scenario where $ \mu_i$ depend on  the
vacuum expectation values of some fields $S$, its fermionic partner 
$\tilde S$ plays the role of a sterile neutrino. This situation is
quiet  generic in the case of models arising from M-theory where
$S$ is a modulus field. Observable  effects are expected to be seen
if the mass and mixing of  $\tilde S$ with active neutrinos  lie in an
interesting region of parameters. This is naturally the case in gauge
mediated supersymmetry breaking and some class of no-scale
supergravity  models where the gravitino mass is very small. For models 
with gravitino mass $m_{3/2}\sim$ TeV, we discuss the possibility 
that the modulino 
mass is  of the order of $m^2_{3/2}/M_{Pl}$. }

\section{Introduction}

Observations of the solar \cite{solar}, atmospheric \cite{atm} anomalies and 
LSND events \cite{lsnd} are experimental hints for
non-zero neutrino mass and  mixing.  It is difficult to
explain simultaneously all of these observations by masses and mixing of
only the three known neutrinos. If the present data are confirmed one may 
need
to introduce a very light fermion $\tilde S$, with a mass $m_s < 10$ eV
\cite{ster}, called``sterile neutrino". This fermion mixes with the
active neutrinos leading to  oscillation patterns that would
explain the experimental data. Several models of the singlet fermions
have been proposed recently \cite{snm}.  In the present work we are 
interested in the
investigation of the possibility that fermionic partners of moduli fields
that appear in most of superstring and M-theory compactifications play the
role of sterile neutrinos. We will argue that the desired neutrino-modulino
mixing appears naturally in models where R-parity is broken through
bilinear terms\cite{BS1}. 

For the particle content of the Minimal Supersymmetric Standard Model the
most general renormalizable superpotential contains an R-parity breaking
part: 
\begin{eqnarray} W_{nR}&=& \lambda_{ijk} L_i L_j E^c_k
+\lambda_{ijk}' L_i Q_j D^c_k
      +\lambda_{ijk}'' D^c_i D^c_j U^c_k \nonumber \\ &+&\mu_i L_i H_2
\end{eqnarray}
Here $m_{e_i},$ $m_{d_i},$ $m_{u_i}$ are the fermion masses while
$v_{1,2}$ are the vacuum expectation values (v.e.v.) of the scalar
components of  the superfields $H_{1,2}$. While theories with R-parity 
conserving superpotential have been
thoroughly investigated, the models where R-parity is broken have
only recently been the focus of more attention. This is because    the
size of the couplings of the trilinear terms
$\lambda_{ijk},\lambda_{ijk}'$ and $\lambda_{ijk}''$ in $W_{nR}$ are
strongly constrained by experiments \cite{hall}. In particular they generate
lepton and baryon number violation leading if combined to dangerous
proton decay as well as other exotic low energy processes. The
experimental bounds on these couplings makes  more
natural to assume that they vanish.  This situation  is
somehow similar with the one encountered  within the R-parity conserving
part of the theory for the $ \mu H_1 H_2$ bilinear term. The size of
the mass parameter  $\mu$ expected  to be of the order of the Planck mass 
is  suppressed and has to be of the order of the electroweak
scale. The appearance of such scale at the tree level of the
fundamental theory, where the only present scale is the Planck scale,
needs a lot of fine-tuning and thus seems very unatural. This 
is the
so-called $\mu$-problem. The most satisfactory solution is to consider
that the $\mu$-term is triggered by supersymmetry breaking. It is thus
naturally of the same magnitude than the soft terms. In a similar way,
if the superpotential $W_{nR}$ is generated at the supersymmetry
breaking scale  with a ``reasonable'' suppression of the different
coefficients, it lies in the experimentally allowed region.  Below we
would like to consider the generation of the $\mu_i L_i H_2$ mixing terms 
and some of their implications for neutrino physics.

\section{Origin of $\bar L H$ terms}

For convenience we  denote $H_1$ also by $L_0$ and we use the greek
 indices $\alpha=0, 1, 2, 3$ while the latin indices take the value
 $i=1,2,3$. The generic bilinear term allowed by the gauge
 interactions is then  $\mu_\alpha L_\alpha H_2$. The existence and the magnitude of these terms depend 
crucially on the mechanism used to break supersymmetry and to communicate it
to the observable sector. We give a quick overview below. More details 
will be given elsewhere \cite{BS}.

\subsection {Giudice-Masiero mechanism}

 In  the same way as for the $ \mu H_1 H_2$ term
Giudice-Masiero mechanism \cite{{giu},{ant}} can be used for generating the 
R-parity breaking bilinear terms \cite{BS1}. The main idea is that the 
Kahler potential contains terms of the form: 
\be {\lambda_\alpha \over
M_{Pl}}\int d\theta^4 zL_\alpha H_2 \ \ \ {\rm and}\ \ \
{\lambda'_\alpha \over M_{Pl}^2}\int d\theta^4 z^2L_\alpha H_2
\label{mu1}
\ee 
where $z$ is a field responsible of the breaking of
supersymmetry $\langle F_z \rangle \sim m_{3/2} M_{Pl}$ so that after 
integration, one
gets the right magnitude for both $\mu$ and  $B\mu$ coefficients. 

In M-theory, the  Kahler potential generically  contains  
terms
$K_{\alpha 2} L_\alpha H_2$  where $K_{\alpha 2}$ are functions of the 
moduli $z$. Supersymmetry breaking by the auxiliary field of one of 
these moduli triggers   $\mu$-terms of the form:
\begin{equation}
\mu_\alpha =  \left( K_{H\bar H}K_{L\bar L}\right) ^{-1/2}  \langle
 m_{3/2} K_{\alpha 2} - {\bar F}^{\bar z}
\partial_z K_{\alpha 2} \rangle . 
\label{mu2}
\end{equation}
The relative size of the $\mu_\alpha$s is governed by the dependence of 
$K_{\alpha 2}$ on $z$. As an example if $z$ is one of the $T$ moduli 
describing the size of a compactification, there is often in the large 
radius 
limit a scaling symmetry. Matter fields might have different charges 
(modular weights) under this symmetry and $T$ might appear with different 
powers, leading to different strengths of the $\mu_\alpha$s. 

\subsection {Higher-weight F-term}

 In the presence of a light singlet $N$ with a Yukawa coupling $\lambda_\alpha N 
L_\alpha H_2$ there is a new contribution to $\mu_\alpha$ \cite{BS1}. This is due to 
higher-weight 
F-terms and is not usually included in the standard two-derivatives 
supergravity \cite{ant}. It leads to a contribution of the form:
\begin{equation}
\mu_\alpha =  -\left( K_{H\bar H}K_{L\bar L}\right) ^{-1/2}  \langle
 \lambda_\alpha {\bar F}^{\bar z}K^{N\bar N}   (f^1_z f^2_N+ 
f^2_z f^1_N)\rangle  
\label{mut}
\end{equation}
where $K^{N\bar N}$ is the inverse metric (coefficient of the kinetic 
term) of the singlet , $f^{(1,2)}$ are two complex functions and $z$ is 
the modulus superfield whose auxiliary component vev breaks supersymmetry. 

Note that this possibility was found to break $R$-parity in models where 
the two-derivative lagrangian seems to respect the symmetry \cite{Nar}. The 
same discussion than for the Giudice-Masiero contribution applies here
concerning the relative strengths of the $\mu_\alpha$.

\subsection{ Superpotential induced $\mu$-terms}

While the previous possibilities for generating $\mu$ terms lead to  values
at most of the order of $m_{3/2}$. In scenarios where the gravitino mass 
is very small  this is 
not a satisfactory option. The only known  solution in this case is to 
appeal to  extra (elementary or composite) singlet(s) $N$ with a 
superpotential: 
\be
 \lambda_\alpha N L_\alpha H_2 + W(N) 
\ee

  Models for $ W(N)$ can be built where the
inclusion of soft terms leads at the electroweak breaking scale to a vev
is for $N$ generating the desired $\mu$ terms. The situation with the 
strength of the resulting $\mu_\alpha$ is analog with the problem of 
fermion masses: $\lambda_\alpha$ might remember that the $L_\alpha$ 
arises from different sectors of the theory (as twists in orbifolds) or 
carry different charges under a new (horizontal) symmetry.

One of the implications of the R-parity violating bilinear terms is that the neutrino-neutralino mixing   
leads after electroweak symmetry breaking to a vev for the
sneutrino. This induces a neutrino mass\cite{hall}:
\begin{eqnarray}
m_{\nu}  \approx     9
\frac{m_Z^2}{M_{\tilde Z}}  \left(\frac{\mu_i}{\mu}\right)^2 \left[
\frac{h_B^2}{16 \pi^2} \log \frac{M_{X}^2}{m_W^2} \right]^2~.  
\label{ntmass}
\end{eqnarray}
where $M_X$ is the scale where the soft terms are universal (where supersymmetry is broken in the hidden sector), $h_B$ is the $b$ quark Yukawa coupling, 
$m_Z$, $m_W$, $M_{\tilde Z}$ are the $Z$, $W$ bosons and Zino masses.
A rough estimation for small $\tan \beta \sim 1$ gives 

\be
m_{\nu_\tau}\sim 3\cdot 10^{-8}\left(\frac{\mu_i}{\mu}\right)^2 \frac{m_Z^2}{M_{\tilde Z}}
\ee
For  $M_{\tilde Z}\sim 300$ GeV, this gives $m_{\nu_\tau}\sim\left(\frac{\mu_i}{\mu}\right)^2 10^4$eV.
Models with $\left(\frac{\mu_i}{\mu}\right) \sim 10^{-2}$ might lead to masses in the eV range for the $\tau$-neutrino.

\section{ Appearance of a sterile-active neutrino mixing }

 Superstring compactifications usually
provide  us (among their  massless states) with massless fields singlets 
under the
standard model gauge symmetry.   These singlets can be divided into
two classes  according to the way they interact with the observable
matter:  

(i) The moduli fields: they couple to the light matter fields only
through non-renormalizable interactions  suppressed by power of
$M_{P}$.  Among these fields are the dilaton $S$,  $T_i$ moduli,
$U_i$ moduli, the continuous Wilson lines, the  blowing-up modes of
orbifolds. Moduli  masses are induced by SUSY  breaking. 

(ii) The non-moduli fields: they can have  renormalizable interactions
with standard matter.  String compactifications often lead to one
anomalous $U(1)_A$ among  several gauge factors  $U(1)'$s and to  a
number  of chiral supermultiplets charged under them.   To satisfy the
$D$ and $F$ flatness conditions,  some of these fields  get large
VEV's.  The resulting symmetry breaking generates a  mass matrix
which may have small or vanishing eigenvalues.  In addition to the
Higgs doublets a singlet $S$ could remain light.

Because of the dependence of the different couplings on $S$,
 one can  write:

\be
\mu_\alpha = m_\alpha{ \langle S \rangle \over M_{Pl}}
\ee

this leads to the effective mixing:

\be {  {\mu_\alpha \langle H_2 \rangle} \over \langle S \rangle
 }{\tilde S} L_\alpha \ee

between the active neutrinos and the fermionic partner ${\tilde S} $ of $S$.

The magnitude of this mixing mass:  
\be
m_{S\nu} ={  {\mu_\alpha \langle H_2 \rangle} \over \langle S \rangle
 }
\ee
 depends on the value of ${ \langle S \rangle}$.

There suppose for example that $\mu_\alpha \sim m_W$  and consider the three natural regions to consider for the value of ${ \langle S 
\rangle}$:

1- For most of the string or M-theory moduli fields who have a
geometrical interpretation as the size of most of the compactification 
scales:

\be M_{GUT} \le \langle S \rangle \le M_{Pl} \ \ \ {\rightarrow } \ \ \
10^{-5} {\rm eV} \le m_{S\nu} \le 10^{-3} {\rm eV}
\label{msnu1}
\ee

2- For the modulus giving the size of the segment in M-theory on $S^1/Z_2$
\cite{witt}: 

\be 10^{12} {\rm GeV} \le \langle S \rangle \le 10^{15} {\rm GeV} \ \ \ 
{\rightarrow } \ \ \ 10^{-2} {\rm eV} \le m_{S\nu} \le 10 {\rm eV}
\label{msnu2}
\ee

3- May be more exotic possibility for moduli, but other singlets leads to
\be \langle S \rangle \sim m_{3/2}  \ \ \
{\rightarrow } \ \ \  m_{S\nu} \sim M_W
\label{msnu3}
\ee
that is not in the range of our interest.

\section{ Singlet fermion mass } 
The supergravity mass matrix 
formula for the relevant fermions from chiral  
supermultiplets has  the form:
\begin{equation}
 M^{\alpha \beta }=
m_{3/2}\langle {\cal{G}}^{\alpha \beta }  
- {\cal{G}}^{\alpha \beta \bar \gamma }{\cal{G}}_{\bar \gamma } 
+ \frac 13{\cal{G}}^\alpha
{\cal{G}}^\beta \rangle ~, 
\label{mass}
\end{equation}
where ${\cal{G}} \equiv  K+\ln \left| w \right|^2$,  
$K$ is  the K\"ahler potential 
and $w$ is the superpotential, 
${\cal{G}}^\alpha \equiv \partial 
{\cal{G}}/\partial \phi ^\alpha$, 
${\cal{G}}^{\bar \gamma} \equiv \partial {\cal{G}}/\partial
{\bar \phi}^{\bar \gamma}$ {\it etc.}. 
The physical mass is obtained by 
dressing (\ref{mass}) with  corresponding wave 
function renormalization factors. 
These  factors are  typically  of order one  and 
we will omit their effects here. 
The gravitino  mass 
is given by $m_{3/2}=\langle e^{K/2}w \rangle $.

\subsection{ Slim gravitinos}

The sterile neutrino  mass is of the order of the gravitino mass. The 
present lower bound on the latter is around $10^{-5}$eV \cite{Zwi}.
Such low masses can be obtained for $p\sim 2$ in a class of no-scale
 supergravity models \cite{EEN} where:
\be
K=-3 ln(z+ \bar z)
\ee
and
\be
f_{ab}=\delta_{ab} e^{iAz^q}
\ee
where $q$ is an integer lead to a relation between the gravitino mass and the gaugino mass $M_{1/2}$:
\be
m_{3/2}\sim M_{Pl} ({ M_{1/2}\over M_{Pl}})^p
\ee
with $p=1/(1-{ 2/3}q)$.
The main idea behind these models is that gaugino  and gravitino masses are
given depend on two different functions $K$ and $f_{ab}$ and can be arranged
to get a large hierarchy between the two scales.
function, as in the case of no-scale models discussed above for values 
of $p\sim 2$.

Also in the case of gauge mediated scenario\cite{gmsb}, the  gravitino mass is  in a range of few eV to the the GeV scale. The modulinos might naturally remain as light as $0.1$ to $10$ eV.

These two cases provide examples where the modulinos (or other  singlets) 
might arise in models from M-theory with masses
and mixing  of the desired sterile neutrino.

\subsection {Fat gravitinos}

The situation is more complicate in the case where $m_{3/2}$ is of the
order of TeV or heavier. A  mass of order $m_{3/2}$ comes from 
interactions  of the form:

\begin{equation}
{\lambda \over M_{P}} \int d^2 \theta z SS
\label{mus}
\end{equation}
 with $z$ the ``goldstino superfield".

A sufficiently light singlet  needs $\lambda$ 
to be very small. A 
particular example where this situation would exists is that of $S$ 
is identified with a twisted state in orbifold compactifications. If $z$ 
is an 
untwisted field  then it was found in that $\lambda$ vanishes. If in 
contrast $z$ and $S$ are both  
twisted with a $Z_2$ or $Z_4$ twist, then one could consider the 
reasonable Kahler potential expansion:
\begin{equation}
K_2=K_{s{\bar s}}{\bar s} s +{z^2 \over M^2_{Pl}}( ss+ h.c.)+...~,
\label{kahb}
\end{equation}
and the superpotential vanishes. Then
according to  (\ref{condi}) a desired mass
of $S$ can be obtained  for
$\langle Z \rangle = \Lambda_{hid}$ and
$\langle {\cal{G}}_{\bar \gamma }\rangle = 0$,
or for $\langle Z \rangle = m_{3/2}$
and  $\langle {\cal{G}}_{\bar \gamma }\rangle \sim 1$.

Diagonalizing mass matrix (\ref{mass}),  
and writing ${\cal G} $ in terms of mass eigenstates  
we get  a necessary condition for the mass of  
the singlet  $S$  to be  of the order $m_{3/2}^2/M_{P}$: 
\begin{equation} 
\left\langle 
{\cal{G}}^{ss} 
- {\cal{G}}^{ss {\bar \gamma} }{\cal{G}}_{\bar \gamma} 
\right\rangle 
\sim {m_{3/2} \over M_{P}}  
\label{condi} 
\end{equation} 
while  typically  it is of order ${\cal O} (1)$ 
\cite{rou}.
Allowing for arbitrary dependence of the superpotential on the moduli, it
is possible to find functions ${\cal{G}}$ (that is, $K$ and $w$) which
satisfy (\ref{condi}).

 As an example consider the case of moduli with large ($\sim
M_P$) VEV's with Kahler
potential of the form: 
 \begin{equation} K = p\ln(\bar s+s ) + ...
\label{kaha} \end{equation} 
where $s \equiv S/M_{P}$;  $p$ is an integer
(typically $p=-1,-2,-3$). If a theory is invariant under shifts $ S
\rightarrow S+i$ \cite{ds}, then the full superpotential would have an
expansion in powers of $e^{-2\pi s}$:  $w \sim e^{-2\pi a s} \sum_n a_n
e^{-2\pi n s}$.

For $ a \sim p/2\pi\left\langle(\bar s+s )
\right\rangle$,  it is easy to find  solutions of (\ref{condi}) with
$s \sim 1$ with ${\cal{G}}_{\bar \gamma} = 0$ or ${\cal{G}}^{ \gamma} \sim 1$ and $\left\langle Z \right\rangle \sim m_{3/2}$.  For other values of $a$ 
(\ref{condi}) implies  large coefficients $a_n \sim e^{2\pi n}$.  Such  large  $a_n$ 
are not excluded. For instance, the  modular 
forms like the $j$-function \cite{tb} which can appear in  
theories with an  $SL(2,{\bf Z})$ symmetry have such 
large coefficients.

Moduli fields usually describe some geometrical
patterns of the compactification they might be subject to some discrete
symmetries. The scalar and fermion singlet components masses depend  on the 
supersymmetry breaking mechanism. The main point is that modulino mass is as induced by (\ref{mus}) depends on the interactions between $S$ and the goldstino direction. Recent developments in M-theory might help understanding the physics of moduli and sup
ersymmetry breaking and thus shade a light on the question of their mass.

When might also try to build models where the smallness of $m_S$ can originate from mixing of $S$ 
with fields , $\phi$, getting a Planck scale mass, if  
the  $S \phi$-mixing  is  the order  
$m_{3/2}$. The latter scale  can appear in the same way as  
the $\mu$-term appears. 
It can be protected by additional $U(1)'$ gauge symmetry   
broken at $m_{3/2}$, if $S$ is charged under $U(1)'$,  
whereas $\phi$ is a singlet of this group. 
Then for the mass of $S$ we 
have the usual see-saw formula: $m_S = m_{3/2}^2/M_{P}$. 

Another possibility is when the  superfield $S$ charged under $U(1)'$ gets 
a VEV of the order $m_{3/2}$. This VEV will lead 
to mixing of the fermion $S$ and gaugino associated with 
$U(1)'$. If this gaugino has the Majorana mass of the 
order $M_P$, then again the see-saw mechanism leads to the desirable mass 
of $S$.

\section { Experimental signature of a sterile neutrino}  

A manifestation of $\tilde S$  depends on  its  mixing angle $\theta$  with 
active neutrinos:
\begin{equation}
\tan 2\theta = \frac{2 m_{\nu S}}{m_S - m_{\nu}}~,   
\label{angle}
\end{equation} 
where $m_{\nu}$ is the neutrino mass and  oscillation parameters, 
$\Delta m^2 \equiv m_S^2 - m_{\nu}^2$.


Consider first scenarios like the no-scale models where $m_{3/2}$ can be of the order of $10^{-3}$eV, or suppressed modulino masses as for  $\sim m^2_{3/2}/M_{Pl}$ in case of a heavy gravitino.  One gets
values $\Delta m^2$ ~$(\approx m_S^2)$ and  $\sin^2 2\theta$ 
in the range of  small mixing solution 
of the $\nu_{\odot}$-problem via the resonance
conversion  $\nu _e\rightarrow {\tilde S}$ in the Sun.

Let us consider the  mixing of ${\tilde S}$  with the electron neutrino $m_{\tilde S} > m_{1}$  ($m_1$ is  the mass of  the dominant component of $\nu_e$). Forthcoming 
experiments, and in particular 
SNO,  will be able to establish whether 
conversion of solar neutrinos to singlet 
state takes place.  

Sensitivity of the $\nu_{\odot}$ data to the neutrino parameters 
is determined by the adiabaticity condition 
for lowest detectable energies ($E \sim 0.2$ MeV):    
\begin{equation}
\Delta m^2 > 
\frac{4 \cdot 10^{-10}~ {\rm eV}^2}{\sin^2 2\theta} 
\label{sens}
\end{equation}
We find  that 
$\nu_{\odot}$ experiments are sensitive to 
$m_{\nu S} > 10^{-5}$ eV.  

If on the contrary the mass and mixing of ${\tilde S}$ are outside  the 
region of solutions of the 
$\nu_{\odot}$
-problem and  some other mechanism 
is responsible for the deficit of 
the  $\nu_{\odot}$-fluxes, then   
the $\nu {\tilde S}$-mixing gives corrections to the main solution. 

Let us  for example assume for  that the  neutrino mass 
spectrum has a  hierarchy with 
$m_2 \sim (2 - 4)\cdot  10^{-3}$ 
eV in the range of solution of    
the $\nu_{\odot}$-problem via 
$\nu _e\rightarrow \nu_{\mu}$ conversion, 
$m_3 \gg m_2$ and $m_1 \ll m_2$.   
There are two generic consequences of the 
$\nu {\tilde S}$-mixing: 
  
(i). Final  neutrino flux  contains not only 
the electron and muon components but also the ${\tilde S}$- component. 
Moreover, the content (relative values of different fluxes) 
depends on neutrino energy.  
Future measurements of the neutral current interactions, 
and in particular, the  ratio 
of neutral to charged current events,  
$(NC/CC)$, in different parts 
of the energy spectrum 
will allow to check  the presence of ${\tilde S}$-flux. 
  
(ii). A dependence of the $\nu_e$-suppression factor 
on energy (so called ``suppression pit") is  
modified. In particular, one may expect 
an appearance of second pit or 
the narrow dip in the non-adiabatic 
 or  adiabatic edges of the  two neutrino suppression pit \cite{BS}. 
This can be revealed in  measurements of  energy 
spectra of the boron- or  $pp$ - neutrinos. 

The system of three states,  
${\tilde S}$, $\nu_e$, $\nu_{\mu}$, relevant for the problem,      
has in general three
resonances. The interplay of transitions in these resonances 
leads to a variety of  possible effects 
which depend on the  adiabaticity conditions in different 
resonances  and  on the  mass of ${\tilde S}$. 
 
If $m_S > m_2$, then system 
has two resonances $\nu_e - {\tilde S}$ and $\nu_e - \nu_{\mu}$. 
Analyzing level crossing scheme we find \cite{BS} that flavor composition 
of the final flux
can change with increase of neutrino energy in the following way: 
$(\nu_e)$ 
$\rightarrow$  
$(\nu_e, \nu_{\mu}~ {\rm or}~  \nu_{\mu})$  
$\rightarrow$
$({\tilde S}~{\rm or}~ \nu_{\mu}, {\tilde S} )$   
$\rightarrow$  
$(\nu_e, \nu_{\mu}, {\tilde S} )$,    
(here dominant components are indicated only). 

If  $m_S <  m_2$  the system has   three  resonances and   
a change of the flavor composition 
with increase of  neutrino energy 
can be as follows: 
$(\nu_e)$   
$\rightarrow$ 
$(\nu_e, {\tilde S}~ {\rm or}~ {\tilde S})$  
$\rightarrow$ 
$({\tilde S}, \nu_{\mu}~{\rm  or} ~ \nu_{\mu})$ 
$\rightarrow$ 
$(\nu_e, \nu_{\mu}, {\tilde S})$   
$\rightarrow$ 
$({\tilde S})$  
$\rightarrow$ 
$(\nu_e, \nu_{\mu}, {\tilde S} )$ .

For $m_S < m_1 < m_{\nu_S}$ the  
$\nu_e {\tilde S}$ mixing  is large,  so that  
vacuum oscillations 
$\nu_e \leftrightarrow {\tilde S}$ on the way from the 
Sun to the Earth become important.  
If $\Delta m^2 \gg 10^{-10}$ eV$^2$,  there is  an 
energy independent suppression 
of the $\nu_e$-flux by 
factor $1 - 0.5 \sin^2 2\theta_{e S}$ 
for  the energies outside 
$\nu_e - \nu_{\mu}$ suppression pit.  
For smaller values of $\Delta m^2$ 
one expects non trivial interplay of the vacuum 
oscillations and  resonance conversion.  
If $m_S < m_{eS} \sim 10^{-5}$ eV,  then 
$\nu_e \leftrightarrow S$ vacuum oscillations alone 
can explain the $\nu_{\odot}$ data.

Let us now  consider other possible consequences of the 
$\nu {\tilde S}$ - mixing. 
Models of  supernovas predict  power dependence 
of density $\rho \propto R^{-3}$ below the envelope,  
in contrast with exponential dependence for the Sun.  
Therefore  the dependence of the  adiabaticity condition 
on the oscillation parameters 
differs from (\ref{sens}),  
and consequently, for 
the border of the sensitivity region we get  
\begin{equation}
\Delta m^2 > A \frac{10^{-8}~{\rm eV}^2}{ \sin^3 2\theta }~.   
\label{sensSN}
\end{equation}
Here  $A \sim {\cal O}(1)$  depends on the model of  star.  
As follows from fig.1,  the $\nu {\tilde S}$- 
mixing can lead to appreciable transitions 
for $\Delta m^2 < 10^{-1}$ eV$^2$. This inequality 
corresponds  
via the resonance condition to densities
$\rho < 10^5$ g/cm$^3$. (For larger 
$\Delta m^2$ and 
larger densities the mixing mass $m_{\nu S}$ can not satisfy the 
adiabaticity condition.)  
Therefore $\nu {\tilde S}$-mixing   
does not influence  dynamics of collapse.  
For this the resonance transition should take place 
at  $\rho > 10^{11}$ g/cm$^3$.  
Also this mixing has no impact on 
the supernova nucleosynthesis which occurs  
in central regions 
where  $\rho > 10^{6}$ g/cm$^3$ \cite{qian}. 
The $\nu {\tilde S}$-mixing can, however,  
lead to a resonance conversion in external regions of 
star thus strongly modifying  properties of neutrino fluxes  
which can be detected on  the Earth.   
For instance, if the neutrinos 
have  the mass hierarchy:  
$m_3 = 1 - 10$ eV, $m_1 \ll m_2 = 10^{-3} - 10^{-1}$ eV  
and  $m_S < m_1$, then the resonance 
conversion $\bar \nu_e \rightarrow \bar S$ 
will  lead to partial or complete disappearance of  the 
$\bar \nu_e$ - signal. The observation of the 
$\bar \nu_e$ signal from SN87A  allows to put a bound 
on  $\bar \nu_e \rightarrow {\tilde S}$ transition \cite{MS}. 
Furthermore, if the 
adiabaticity condition is fulfilled  in 
$\nu_{\mu} {\tilde S}$-resonance, then conversions  
$\nu_e \rightarrow \nu_{\tau}$ and $\nu_{\mu} \rightarrow {\tilde S}$ 
lead also to  disappearance of $\nu_e$-flux. 
Notice that without $\nu_{\mu}{\tilde S} $-mixing  one would  expect 
$\nu_{\mu} \rightarrow \nu_e$ transition 
which results in a hard (corresponding to 
initial $\nu_{\mu}$) $\nu_e$-spectrum.

For  $m_1 < m_S < m_2$ the resonance 
conversion $\nu_e \rightarrow {\tilde S}$   
occurs in supernovas. A simultaneous transition 
$\nu_{\mu} \rightarrow \nu_e$ will lead 
to hard $\nu_e$-spectrum and disappearance of the 
$\nu_{\mu}$-flux.

A different possibility is that all neutrinos are massless, then 
$\nu \rightarrow {\tilde S}$ transition 
can solve the $\nu_{\odot}$-problem. For supernova 
neutrinos it can lead   
to disappearance of the $\nu_e$-flux  
and to conversion $\bar{\nu}_{\mu} \rightarrow \bar {\tilde S}$. 

The $\nu - {\tilde S}$ oscillations in the Early 
Universe generate ${\tilde S}$ components 
which increases the expansion rate of the Universe 
and therefore influences the 
primordial   nucleosynthesis \cite{NS}. 
The  $\nu {\tilde S}$-mixing is important  
for $\Delta m^2 < 10^{-1}$ eV$^2$.


Scenario like the gauge mediated supersymmetry breaking (and some class of no-scale models) may prefer a modulino mass ( roughly of the order of the gravitino mass) in the range 0.1-10 eV. The natural mixing mass  $m_{S\nu}$ lies between $10^{-3}m_S$ and
 $m_S$. Phenomenological implications of this scenario are under investigation.

\section { Conclusion}
In conclusion, M-theory (or string) compactification 
often leads to moduli superfields, singlet
 of standard model group, which couple with observable matter 
via the Planck mass  suppressed constant. 
If some of them remain very light as suggested in this work then 
they will have 
a number of manifestations in neutrino physics. 
Forthcoming experiments will be able to check 
whether neutrinos have mixing with such singlets.

\section*{Acknowledgments}
This talk is based  on work  with  A. Yu. Smirnov,  part of it published\cite{BS1}, with   whom I enjoyed collaborating. I wish to thank the ICTP for their warm hospitality during  PAST97 conference and the Extended Workshop  Highlights in Astroparticle P
hysics. My work is supported  by the DOE grant DE-FG03-95ER40917.

\end{document}